# A Video Coding Method Based on Neural Network for CLIC2024

Zhengang Li, Jingchi Zhang, Yonghua Wang, Xing Zeng, Zhen Zhang,
Yunlin Long, Menghu Jia, Ning Wang

State Key Laboratory of Mobile Network and Mobile Multimedia Technology

ZTE Corporation

{li.zhengang1, zhang.jingchi, wang.yonghua1, zeng.xing1, zhang.zhen23,
long.yunlin, jia.menghu, wangning}@zte.com.cn

*Abstract*: This paper presents a video coding scheme that combines traditional optimization methods with deep learning methods based on the Enhanced Compression Model (ECM). In this paper, the traditional optimization methods adaptively adjust the quantization parameter (QP). The key frame QP offset is set according to the video content characteristics, and the coding tree unit (CTU) level QP of all frames is also adjusted according to the spatial-temporal perception information. Block importance mapping technology (BIM) is also introduced, which adjusts the QP according to the block importance. Meanwhile, the deep learning methods propose a convolutional neural network-based loop filter (CNNLF), which is turned on/off based on the rate-distortion optimization at the CTU and frame level. Besides, intra-prediction using neural networks (NN-intra) is proposed to further improve compression quality, where 8 neural networks are used for predicting blocks of different sizes. The experimental results show that compared with ECM-3.0, the proposed traditional methods and adding deep learning methods improve the PSNR by 0.54 dB and 1 dB at 0.05Mbps, respectively; 0.38 dB and 0.7[1]dB at 0.5 Mbps, respectively, which proves the superiority of our method.

## 1. Introduction

Recent years have witnessed the rapid development of video compression applications. The video data volume increases rapidly, which constantly brings new challenges to video coding. Many video coding standards have been released such as Versatile Video Coding (VVC)[1] , and High Efficiency Video Coding (HEVC)[2], which greatly promote the development of video compression techniques and related industry development.

Nowadays, deep-learning-based coding tools have attracted a lot of attention in exploration experiments, which mainly concentrate on modifying the prediction and filtering modules, which bring significant performance improvement in terms of

---

objective indicators and subjective quality. Meanwhile, they also bring great complexity to encoding and decoding, which brings great difficulty to practical applications.

In this paper, we choose the ECM[3] as the baseline that could bring more bit-rate savings than the VVC reference software VTM[4]. To further enhance the quality and reduce the compression artifacts of compressed frames, CNNLF and NN-intra are proposed based on ECM. In addition, we also have optimized traditional encoding methods including QP adaptive adjustment and Block importance mapping.

The remainder of this paper is organized as follows: traditional encoding method optimization will be introduced in Section 2, CNNLF will be described in Section 3, and NN-intra in Section 4. Experimental results will be presented and analyzed in Section 5 and the conclusion will be given in Section 6.

## 2. Optimization of traditional encoding

The goal of CLIC is to maximize subjective scores, so a perceptual QP adaptation algorithm targeting subjective effect maximization should be applied[5]. For the best encoding performance, a Group Of Pictures (GOP) size of 32 pictures[6] is recommended in the JVET common test conditions in RA configuration. Additionally, using only one key frame at the beginning of the sequence saves bit allocation.

### 2.1. QP adaptive adjustment

By analyzing the content characteristics of video sequences, static or slowly changing sequences can be distinguished from fast-moving sequences. For slow-moving video sequences, the better the encoding quality of the I frame, the better the encoding quality of the overall sequence. Therefore, it is very important to set the appropriate first frame QP. Here, the initial I frame QP is set through rate control method, and then perceptual quantization coding is performed[7]. After testing, the final selection range of the QP offset value at 0.05Mbps is shown in Figure 1. The abscissa represents the negative QP offset value, and the ordinate represents the number of sequences in the verification set. For I frame, this article only uses the above method to set the QP offset value for slowly changing sequences, as shown in the blue column in Figure 1, the original QP offset value -3 is used for fast-moving sequences, as shown in the green column in Figure 1.

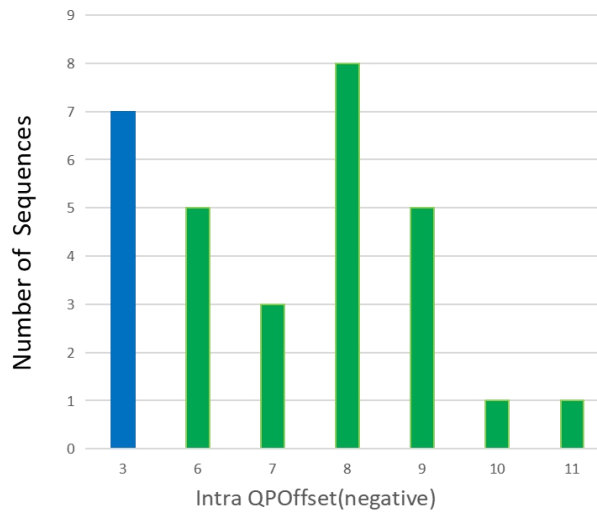

**Figure 1:** Selected QP Offset values

## 2.2. Block importance mapping

Block importance mapping is a new algorithm that estimates the estimated importance of a given CTU for future pictures and obtains a delta QP value for each CTU. The selected QP is in the range of -2 to +2 relative to the CTU QP after perceptual quantization adjustment[8].

The importance of a given CTU is estimated by measuring the difference between motion-compensated pictures, which is based on calculations performed in the temporal filter. QP is only modified for frames where the temporal filter is active, i.e., where picture order count (POC) is divisible by eight.

After motion compensation for each 16x16 block, the temporal filter computes the variance of the original block and sum of squared difference (*SSD*), the sum of squared differences between the original block and the corresponding block in the reference picture after motion compensation. Based on this, an error E is computed as

$$E = 0.2 \cdot \frac{SSD+5}{V+5} + \frac{SSD}{3200} \qquad (1)$$

The E-value is currently computed per 16x16 block of each reference picture in the temporal filter. For this method, the average E-value in a CTU is calculated for the pictures immediately before and after the current frame. These two values are then averaged to form a value E1 for each CTU in the current frame, and the process is repeated for both frames at a distance of two to generate *E2* for each CTU. The derivation gives the E3 value for each CTU:

$$E3 = max(E1, E2) + abs(E2 - E1) * 3 \qquad (2)$$

The E3 value takes into account the speed at which the importance disappears. If E1 indicates high importance and E2 indicates low importance, the importance disappears quickly, indicating that the block is not that important. However, if both E1 and E2 indicate high importance, it indicates that the importance remains longer and good encoding of the current block is more important.

Finally, a threshold is performed to decide a delta QP for each CTU, as shown in Table 1.

**Table 1:** Table of corresponding values delta QP of E3

| E3 | 0-22 | 23-41 | 42-76 | 77-101 | >102 |
|---|---|---|---|---|---|
| delta QP | -2 | -1 | 0 | +1 | +2 |

## 3. Adaptive In-Loop Filter based on CNN

Main operations used in video codecs such as intra prediction, inter prediction, and quantization are performed block by block. Therefore, the coding parameters vary from block to block, resulting in blocking effects. In addition, high-frequency components of the video will be lost during the quantization process, which leads to ringing and blurring effects.

Aiming at eliminating these compression artifacts, an adaptive in-loop filter based on neural network is adopted. The CNNLF is integrated into ECM-3.0 to serve as an in-loop processing module for better compression quality. After CNNLF processing, on the one hand, the image quality of this frame can be improved, on the other hand, it can provide a better reference for other frames. The architecture of network and training process will be introduced separately.

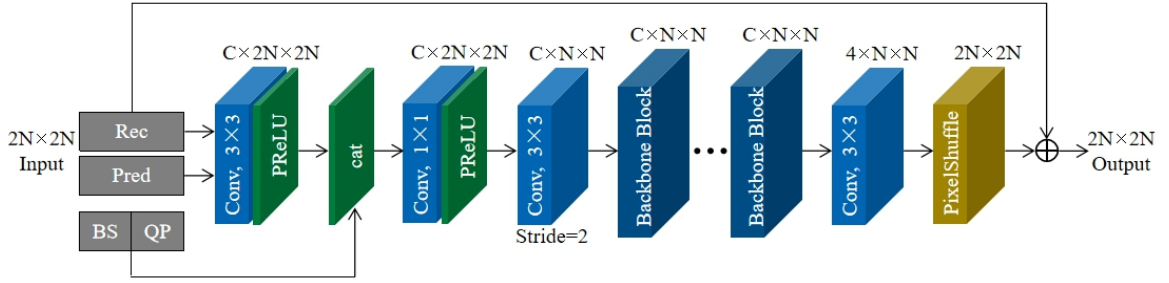

**Figure 2(a):** CNNLF architecture

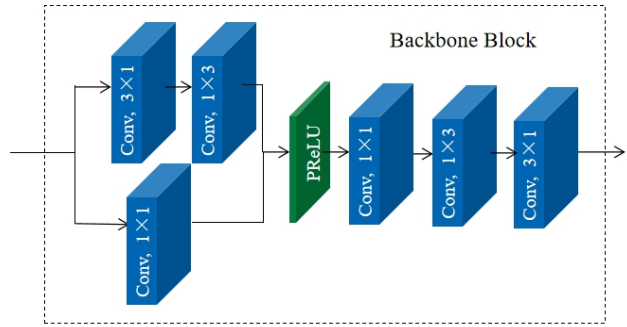

**Figure 2(b):** Backbone Block

## 3.1. Architecture of Network

The proposed CNNLF structure diagram is shown in Figure 2(a). The input of the network contains reconstructed image (Rec), predicted image (Pred), boundary strengths (BS), and quantization parameter (QP). C denotes the channel number of feature maps, which is set as 64. N represents the size of patch size in one dimension, and the number of backbone blocks is set as 8[9].

CNNLF is conditioned on QP, leading to a unified model to handle different quality levels. Based on conventional residual blocks, the backbone block introduces some optimization techniques: 3x3 convolution CP decomposition, wide activation, and increasing receptive field, as shown in Figure 2(b).

## 3.2. Training Process

For the Network, PyTorch is used as the training platform. For the training data, we randomly extracted and cropped consecutive frames from the CLIC 2024 validation data set and BVIDVC data set[10] about 200k samples. Then, we convert the data set to YUV420 format and use ECM-3.0 for encoding and decoding. In order to make CNNLF obtain stronger generalization ability, configure RA for B-slice training and AI for I-slice training. We trained for 90 epochs using the Adam optimizer with a starting learning rate of 1e−4, and a batch size of 64, and reconstructed images were split into 128x128 luma and 64x64 chroma blocks. We decay the learning rate by 0.1 whenever the loss reaches a plateau. The network is first optimized for MSE to improve convergence and stability, then switches to the SSIM metric at 80% of the total training steps, and also increases the patch size to reduce border artifacts.

## 4. Intra prediction using neural networks

### 4.1. Architecture of neural networks

The neural network-based intra-prediction mode contains 8 neural networks, each predicting blocks of a different size in $\{4 \times 4, 8 \times 4, 16 \times 4, 32 \times 4, 8 \times 8, 16 \times 8, 16 \times 16, 32 \times 32\}$. The neural network $f_{h,w}(., \theta_{h,w})$ predicting $w \times h$ blocks is convolutional, where $\theta_{h,w}$ gathers its parameters. For a given $w \times h$ block Y, $f_{h,w}$ takes the context $X$ made of $n_a$ rows of $n_l + 2w$ reference samples located above this block and $n_l$ columns of $2h$ reference samples on its left side to provide Ŷ. Besides, $f_{h,w}(., \theta_{h,w})$ gives two indices grpIdx$_1$ and grpIdx$_2$, grpIdx$_i$ denotes the index characterizing the LFNST kernel index. The "flattening" operation flattens its input stack of feature maps, the "concatenation" operation concatenates its two input vectors into a single vector, as in Figure 3[11].

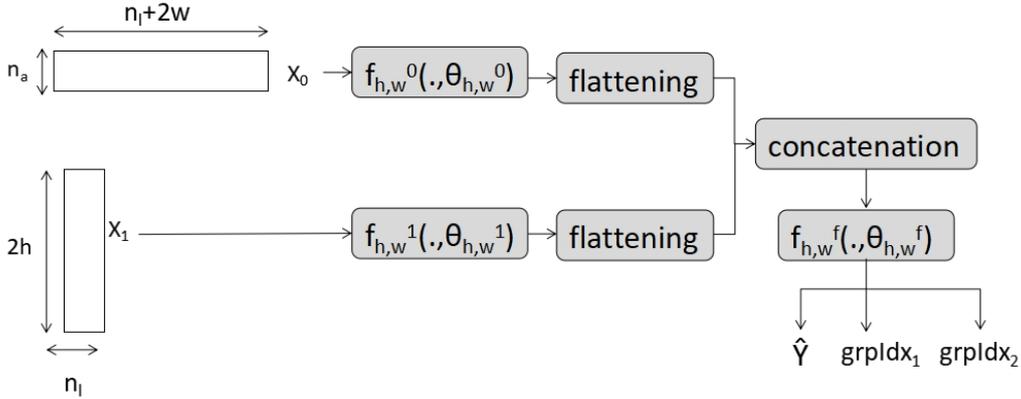

**Figure3:** prediction of the current $w \times h$ block $Y$ from its context $X$ of neighboring reference samples via the neural network $f_{h,w}(., \theta_{h,w})$

### 4.2. Training Process

Similar to section 3.2, there are some differences: the NN-intra models are trained separately to adapt to different QP points, and QP points include 27, 32, 37, 43. The configuration for AI in ECM-3.0 is used to generate the compressed training data. We trained for 15 epochs using the Adam optimizer with a starting learning rate of 1e−4, a batch size of 64, and $f_{h,w}(., \theta_{h,w})$ is trained on pairs of a $w \times h$ block and its context.

## 5. Experimental Results

### 5.1. Implementation

The proposed traditional methods, CNNLF and NN-intra are implemented on top of the ECM-3.0 reference software. Deblocking filtering and SAO are disabled while ALF is placed after the proposed CNNLF. During the process of CNNLF, whether to apply the proposed filter is based on the rate-distortion optimization in CTU and frame level. In addition to CNNLF with CTU and frame-level flags, the flag for on/off CNNLF is also in the encoder conditional parameter index.

In order to compare the performance with ECM-3.0, we adopt the same coding parameters under the default configuration of RA, and perceptual QP adaption is enabled. Then after converting the mp4 videos in the validation set to YUV videos, we encoded them at 0.05 Mbps and 0.5 Mbps, respectively. Finally, the objective coding performance index PSNR and human visual subjective effects are compared.

### 5.2. Compression Performance

The target bitrate is approximately 0.05 Mbps and 0.5 Mbps for the 30 sequences of 10 second. Consequently, the limit for the Submission Size is set to 18.9877 Mbytes at 0.5 Mbps and 1.8988 Mbytes at 0.05 Mbps, respectively[12].

**Table2:** The compression performance of in the validation set of CLIC at 0.05 Mbps

| Method | Data Size (Mbytes) | PSNR (dB) |
|---|---|---|
| ECM-3.0 | 1.75 | 27.26 |
| Trad_Opt | 1.78 | 27.80 |
| Trad_Opt+CNNLF+NN-intra | 1.79 | 28.26 |

**Table3:** The compression performance of in the validation set of CLIC at 0.5Mbps

| Method | Data Size (Mbytes) | PSNR (dB) |
|---|---|---|
| ECM-3.0 | 18.82 | 34.22 |
| Trad_Opt | 18.75 | 34.60 |
| Trad_Opt+CNNLF+NN-intra | 18.68 | 34.92 |

Experimental results demonstrate that the proposed video compression approach can achieve great performance in the validation sets of CLIC 2024 under the condition of limiting the bitstream size. On objective indicators, for 0.05 Mbps, as shown in Table 2, the proposed traditional optimization method, named Trad_Opt, improves PSNR by 0.54 dB, and adding CNNLF and NN-intra improves PSNR by 1 dB. And for 0.5 Mbps, as shown in Table 3, Trad_Opt improves PSNR by 0.38 dB, adding CNNLF and NN-intra improves PSNR by 0.7 dB. In subjective evaluation, the compression artifacts were greatly reduced and the visual effects were significantly improved.

In conclusion, the proposed method not only has a suitable amount of data but also has a better objective and subjective performance than ECM-3.0, which strongly proves the superiority of our method.

## 6. Conclusion

In this paper, A hybrid video codec based on ECM was proposed. QP adaptive adjustment and BIM technology were introduced in this codec. As for the deep learning solution, CNNLF and NN-intra methods were proposed to further improve compression quality. The CNNLF network architecture mainly consists of the backbone block module and the convolution feature map module. On the validation set of the Challenge for Learning Image Compression (CLIC), the proposed method achieves more remarkable compression quality than the conventional ECM 3.0 in terms of PSNR and subjective evaluation.